\documentclass[a4paper]{article}
\usepackage{color}
\usepackage{orcidlink}
\usepackage{mathtools}
\usepackage{paralist}
\usepackage{listings}
\usepackage[backend=bibtex]{biblatex}
\bibliography{CompCert_TCB}
\usepackage{hyperref}
\usepackage[hyphenbreaks]{breakurl}

\newcommand{\keywords}[1]{}

\newcommand{\code}[1]{{\small\texttt{#1}}}

\lstdefinelanguage{Coq}{ 
mathescape=true,
texcl=false, 
morekeywords=[1]{Section, Module, End, Require, Import, Export,
  Variable, Variables, Parameter, Parameters, Axiom, Hypothesis,
  Hypotheses, Notation, Local, Tactic, Reserved, Scope, Open, Close,
  Solve, Obligations,
  Delimit, Extract, Inlined, Constant,
  Definition, Let, Ltac, Fixpoint, CoFixpoint,
  Morphism, Relation, Implicit, Arguments, Unset, Contextual,
  Strict, Prenex, Implicits, Inductive, CoInductive, Record,
  Structure, Canonical, Coercion, Context, Class, Global, Instance,
  Program, Infix, Theorem, Lemma, Corollary, Proposition, Fact,
  Remark, Example, Proof, Goal, Save, Qed, Defined, Hint, Resolve,
  Rewrite, View, Search, Show, Print, Printing, All, Eval, Check,
  Projections, inside, outside, Def},
morekeywords=[2]{forall, exists, 
exists2, fun, fix, cofix, struct,
  match, with, end, as, in, return, let, if, is, then, else, for, of,
  nosimpl, when},
morekeywords=[3]{Type, Prop, Set},
morecomment=[s]{(*}{*)},
showstringspaces=false,
morestring=[b]",
morestring=[d],
tabsize=3,
extendedchars=false,
sensitive=true,
breaklines=false,
basicstyle={\ttfamily\small},
captionpos=b,
columns=[c]fixed,
frame=single,
identifierstyle={\ttfamily},
keywordstyle=[1]{\ttfamily\bfseries\color{magenta}},
keywordstyle=[2]{\ttfamily\bfseries\color{blue}},
keywordstyle=[3]{\ttfamily\bfseries},
stringstyle=\ttfamily,
commentstyle={\ttfamily\color{red}\itshape},
literate=
    {:=}{{$\coloneqq$}}1
    {forall}{{$\forall$}}1
    {exists}{{$\exists$}}1
    {<-}{{$\leftarrow$}}1
    {=>}{{$\Rightarrow$}}2
    {<=}{{$\!\leq\!$}}1
    {>=}{{$\!\geq\!$}}1
    {==}{{\code{==}}}1
    {==>}{{\code{==>}}}1
    {->}{{$\rightarrow$}}1
    {~>}{{$\leadsto$}}1
    {<~}{{$\leftsquigarrow$}}1
    {<->}{{$\leftrightarrow\;$}}1
    {\/\\}{{$\wedge$}}1
    {\\\/}{{$\vee$}}1
    {++}{{\code{++}}}1
    {...}{{${\ldots}$}}1
}[keywords,comments,strings]

\lstnewenvironment{lstcoq}{\lstset{language=Coq}}{}

\title{The Trusted Computing Base of the CompCert Verified Compiler\footnote{This is an erratum version of our ESOP'22 paper that fixes footnote~\ref{foot:fixoct2022}.}}

\author{David Monniaux\orcidlink{0000-0001-7671-6126} \and
  Sylvain Boulmé\orcidlink{0000-0002-9501-9606}}

\newcommand{\software}[1]{\textsf{#1}}
\newcommand{\coq}{\software{Coq}}
\newcommand{\ocaml}{\software{OCaml}}
\newcommand{\gcc}{\software{gcc}}
\newcommand{\clang}{\software{clang}}

\newcommand{\CompCert}{\software{CompCert}}
\newcommand{\CompCertS}{\software{CompCertS}}
\newcommand{\CompCertELF}{\software{CompCertELF}}
\newcommand{\CompCertKVX}{\software{CompCert-KVX}}
\newcommand{\CompCertSSA}{\software{CompCert-SSA}}

\newcommand{\modname}[1]{\mbox{\texttt{#1}}}
\newcommand{\progname}[1]{\mbox{\texttt{#1}}}
\newcommand{\compcertcommit}[2]{\href{https://github.com/AbsInt/CompCert/commit/#2}{\textsf{#1}}}
\newcommand{\compcertkvxcommit}[2]{\href{https://gricad-gitlab.univ-grenoble-alpes.fr/certicompil/compcert-kvx/-/commit/#2}{\textsf{#1}}}
\newcommand{\compcertissue}[1]{\href{https://github.com/AbsInt/CompCert/issues/#1}{\textsf{#1}}}

\begin{document}
\maketitle 
\begin{abstract}
  {\CompCert} is the first realistic formally verified compiler: it provides a machine-checked mathematical proof that the code it generates matches the source code.
  Yet, there could be loopholes in this approach. We comprehensively analyze aspects of {\CompCert} where errors could lead to incorrect code being generated.
  Possible issues range from the modeling of the source and the target languages to some techniques used to call external algorithms from within the compiler.
\keywords{Formally Verified Software \and The {\coq} Proof Assistant}
\end{abstract}

\section{Introduction}

{\CompCert}~\cite{Leroy-backend,Leroy_CACM09,CompCert_manual3.9} is a formally verified compiler for a large subset of the C99 language (extended with some C11 features): there is a proof, checked by a proof assistant, that if the compiler succeeded in compiling a C program and that program executes with no undefined behavior, then the assembly code produced executes correctly with the same observable behavior.
Yet, this impressive claim comes with some caveats; in fact, there have been bugs in {\CompCert}, some of which could result in incorrect code being produced without warning~\cite{YangCER11}. How is this possible?

The question of the Trusted Computing Base (TCB) of {\CompCert} has been alluded to in general overviews of {\CompCert}~\cite{Leroy-BKSPF-2016,Kastner-LBSSF-2017}, but there has been so far no detailed technical discussion of that topic.
While our discussion will focus on {\CompCert} and {\coq}, we expect that much of the general ideas and insights will apply to similar projects and other proof assistants: other verified compilers, verified static analysis tools, verified solvers, etc.

We analyze the TCB of the official releases of {\CompCert},%
\footnote{\url{https://github.com/AbsInt/CompCert}}
and two forks: {\CompCertKVX},%
\footnote{\url{https://gricad-gitlab.univ-grenoble-alpes.fr/certicompil/compcert-kvx}}
adding various optimizations and a backend for the Kalray~KVX VLIW (very large instruction word) core, and {\CompCertSSA},%
\footnote{\url{https://gitlab.inria.fr/compcertssa/compcertssa}}
adding optimizations based on single static assignment (SSA) form~\cite{DBLP:conf/esop/BartheDP12,DBLP:phd/hal/Demange12a}.
Versions and changes to these software packages are referred to by git commit hashes.
We discuss alternate solutions, some of which already implemented in other projects, their applicability to {\CompCert}, as well as related work.

Sections~\ref{sec:Coq} and~\ref{sec:axioms} analyze the TCB part coming from {\coq} usage. Section~\ref{sec:frontend} presents the TCB part connecting the {\coq} specification of {\CompCert}'s inputs (source code) to the user view of these inputs. Sections~\ref{sec:assembly} and~\ref{sec:ABI} analyze the TCB part connecting the {\coq} specification of {\CompCert}'s generated programs to the actual platform running these programs.
The conclusion (\ref{sec:conclusion}) summarizes which TCB parts of {\CompCert} (and its forks) are the most error-prone, and discusses possible improvements.

\section{The Coq Proof Assistant}\label{sec:Coq}
{\CompCert} is mostly implemented in {\coq},%
\footnote{\url{https://coq.inria.fr/}} an interactive proof assistant~\cite{Coq_manual}.
{\coq} is based on a strict functional programming language, \emph{Gallina}, based on the Calculus of Inductive Constructions, a higher-order $\lambda$-calculus.
This language allows writing executable programs, theorem statements about these programs, and proofs of these theorems.
{\CompCert} is not directly executed within {\coq}. Instead, the {\coq} code is \emph{extracted} to {\ocaml} code, then linked with some manually written {\ocaml} code.
We now discuss how issues in the {\coq} implementation may impact the correctness of {\CompCert}. 

\subsection{Issues in Coq Proof Checking}\label{sec:CoqKernel}
Proofs written directly in Gallina would be extremely tedious and unmaintainable, so proofs are usually built using {\coq} tactics.
While some other proof assistants trust tactics to apply only correct logical steps, this is not the case with {\coq}: what the tactics build is a $\lambda$-term, which could have been typed directly in Gallina if not for the tedium, and this $\lambda$-term is checked to be correctly typed by the {\coq} kernel.
This allows tactics to be implemented in arbitrary ways, including calling external tools, without increasing the~TCB.

A theorem statement is proved when a $\lambda$-term is shown to have the type of that statement (the Curry-Howard correspondence thus identifies statements and types, and proofs and $\lambda$-terms).
Thus, all logical reasoning in {\coq} relies on the correctness of the {\coq} kernel, and some driver routines.
In addition to the {\coq} compiler \progname{coqc} and {\coq} toplevel \progname{coqtop}, a proof checker \progname{coqchk} provides some level of independent checking.

{\coq} is a mature development, however ``\emph{on average, one critical bug has been found every year in {\coq}}''~\cite{DBLP:journals/pacmpl/SozeauBFTW20}. Let us comment on the official list of these bugs.%
\footnote{\url{https://github.com/coq/coq/blob/master/dev/doc/critical-bugs}}
Interestingly, the list classifies their risk according to whether they can be exploited by accident.
We can probably assume that the designers of {\CompCert} would not deliberately write code meant to trigger a specific bug in {\coq} and prove false facts about compiled code:
exploiting a {\coq} bug by mistake in a way sufficiently innocuous to evade inspection of the source code, to accept an incorrect optimization that would be triggered only in very specific cases (to evade being found through testing), seems highly unlikely.

Proofs are checked by {\coq}'s kernel, which is essentially a type-checker for the $\lambda$-calculus implemented by {\coq} (the Calculus of Inductive Constructions with universes). There have been a number of critical bugs involving {\coq}'s kernel, particularly the checking of the guard conditions (whether some inductively defined function truly performs structural induction) and of the universe conditions ({\coq} has a countable infinity of type universes, all syntactically called \lstinline|Type|, distinguished by arithmetic constraints, which must then be checked for validity).
These conditions prevent building some terms having paradoxical types.
Furthermore, there are options (in the source code or the command-line) that disable checking guard, universe or positivity conditions. For instance, if one disables the guard condition to build a nonterminating function as though it were a terminating one, it is possible to prove ``false'':
\begin{lstlisting}[language=Coq]
Unset Guard Checking.
Fixpoint loop {A: Type} (n : nat) {struct n}: A := loop n.
Lemma false: False.          Proof. apply loop. exact O. Qed.
\end{lstlisting}
\progname{coqchk -o} lists which guard conditions have been disabled---none in {\CompCert}.

The {\coq} kernel can evaluate terms (reduce them to a normal form), but is rather slow in doing so. For faster evaluation, it has been extended with a virtual machine (\progname{vm\_compute})~\cite{DBLP:conf/icfp/GregoireL02} and a native evaluator (\progname{native\_compute})~\cite{DBLP:conf/cpp/BoespflugDG11}. Both are complex machinery, and a number of critical bugs have been found in them.%
\footnote{For instance, there used to be a bug with respect to types with more than 255 constructors that allowed proving ``false'' \url{https://github.com/clarus/falso}, so ludicrous that it made it into a satirical site \url{https://inutile.club/estatis/falso/}.} 
In {\CompCert}, there is a few direct calls to \progname{vm\_compute}, none to \progname{native\_compute}; but there may be indirect calls through tactics calling these evaluators.

\subsection{Issues in Coq Extraction}\label{sec:extraction}
\lstset{language=[Objective]Caml}
{\coq}'s extractor, as used in {\CompCert}, produces {\ocaml} code from {\coq} code, which is then compiled and linked together with some other {\ocaml} code.
Extraction~\cite{Letouzey-CiE-08,DBLP:phd/hal/Letouzey04}, roughly speaking, corresponds to removing non-computational (proof) content, compensating for some typing issues (see below), renaming some identifiers (due to different reserved words), and of course printing out the result.
{\coq}'s extractor and {\ocaml} are in the TCB of {\CompCert}.

{\ocaml}'s type safety ensures that, barring the use of certain features that circumvent this type safety (unsafe array accesses, marshaling, calls to external C functions, the \lstinline|Obj| module allowing unsafe low-level memory accesses\dots), no type mismatch or memory corruption can happen at runtime within that {\ocaml} code.
None of these features are used within {\CompCert}, except for calling C functions implementing the {\ocaml} standard library, and some calls to \lstinline|Obj.magic|, a universal unsafe cast operator, produced by {\coq}'s extractor.

Calls to \lstinline|Obj.magic| are used by the extractor to force {\ocaml} to accept constructs (dependent types, arbitrary type polymorphism) that are correctly typed inside {\coq} but that, when mapped to {\ocaml} types, result in ill-typed programs.
The following program is correct in {\coq} (or in System~F) but cannot be typed within {\ocaml}'s Hindley-Milner style of polymorphism, so uses \lstinline|Obj.magic|:%
\footnote{Some System~F-like polymorphism was added to {\ocaml}: structure types with polymorphic fields. This is not used by {\coq}'s extractor as of {\coq}~8.13.2.}
\begin{lstlisting}[language=Coq,firstline=2]
Definition m (g : forall {T}, list T -> list T) : Type :=
  ((g (false :: nil)), (g (0 :: nil))).     Extraction m.
\end{lstlisting}
The following program, which is similar to some code in the \modname{Builtins0.v} {\CompCert} module, uses dependent types
\begin{lstlisting}[language=Coq]
Inductive data := DNat : nat -> data | DBool : bool -> data.
Definition get_type (d : data) : Type :=
  match d with DNat _ => nat | DBool _ => bool end.
Definition extract (d : data) : get_type d :=
  match d with DNat n => n   | DBool b => b    end.
Require Extraction.  Extraction extract.
\end{lstlisting}
Its extraction uses \lstinline|Obj.magic|:%
\footnote{Variants of this example correspond to general algebratic data types (GADTs), another recent addition to {\ocaml}'s type system not yet exploited by the extractor.}%
\begin{lstlisting}[language={[Objective]Caml}]
let extract = function DNat n  -> Obj.magic n
                     | DBool b -> Obj.magic b
\end{lstlisting}

Thus, incorrect behavior in the {\coq} extractor could, in theory at least, produce {\ocaml} code that would not be type-safe, in addition to producing code not matching the {\coq} behavior.
Is this serious cause for concern?
On the one hand, the extraction process is quite syntactic and generic. It seems unlikely that it could produce valid {\ocaml} code that would compile, pass tests, yet occasionally would have subtly incorrect behavior.%
\footnote{\href{https://github.com/coq/coq/labels/part\%3A\%20extraction}{{\coq}'s bug tracker} lists extractor bugs that, to the best of our knowledge, result in programs that are rejected by {\ocaml} compilers.}
    On the other hand, {\CompCert} is perhaps the only major project using the extractor, which is thus not thoroughly tested.
We do not know of any extractor bug that could result in {\CompCert} miscompiling.
Another related potential source of bugs comes from the link of {\ocaml} code extracted from {\coq} and ``external'' {\ocaml} code. 
This is discussed in Section~\ref{sec:mismatch}.

Sozeau-et-al~\cite{DBLP:journals/pacmpl/SozeauBFTW20} study an approach to reduce the TCB of {\coq} by providing a formally verified (in \coq) implementation of a significant subset of its kernel and paving the road for a formally verified extraction. However, the target language of the extraction ({\ocaml}~?) would still be in the TCB.
An alternative solution would be direct generation of assembly code from Gallina, as done by {\OE}uf~\cite{10.1145/3167089}; however parts of {\CompCert} are currently written in {\ocaml} and would have to be rewritten into Gallina.
{\OE}uf extracts Gallina to Cminor, one of the early intermediate languages of {\CompCert}, then produces code using {\CompCert}.%
\footnote{Other systems meant to generate code from definitions in a proof assistant, generate code directly rather than reuse an existant compiler. This approach is promoted~\cite{DBLP:conf/itp/KumarMTM18} with the argument that such a process is safer than textual extraction to, say, {\ocaml}. This is not so clear to us. On the one hand, extracting (without proof of correctness) Gallina to a subset of {\ocaml}, printing the result, then running the {\ocaml} compiler, surely adds a lot to the TCB. On the other hand, it is typically difficult to get right in a compiler the modeling of the assembly instructions, the ABI, the foreign function interface, as discussed in Section~\ref{sec:assembly}. Bugs at that level are caught by extensive testing. Surely, the {\ocaml} code generator, the many libraries using {\ocaml}'s foreign function interface, are more thoroughly tested by usage than a code generator used to extract a few specific projects developed in a proof assistant.}
CertiCoq%
\footnote{\url{https://github.com/CertiCoq/certicoq}}~%
\cite{DBLP:journals/pacmpl/Paraskevopoulou21,Paraskevopoulou_PhD} also extracts to Clight, which may be compiled with any C compiler.

\section{Use of Axioms in Coq}\label{sec:axioms}
{\coq}, as other proof assistants, checks that theorems are properly deduced from a (possibly empty) set of axioms.
Axioms are also introduced as a mechanism to link Gallina programs to external {\ocaml} code through extraction.
Improper use of axioms may lead to two forms of inconsistency: logical inconsistency and inconsistency between the {\coq} proof and the {\ocaml} external code.

\subsection{Logical Inconsistency}\label{sec:logicalcontradiction}
{\coq} is based on type theory, with logical statements seen through the Curry-Howard correspondence: a proof of a logical statement is the same thing as a program having a certain type. In other words, a theorem is proved if and only if there is a $\lambda$-term inhabiting the type corresponding to the statement of the theorem.
An axiom is thus just the statement that a certain constant, given without definition, inhabits a certain type.

The danger of using axioms is that they may introduce inconsistency, that is, being able to prove a contradiction; from which, through \emph{ex falso quodlibet}, any arbitrary statement is provable. Furthermore, it is possible that several axioms are innocuous individually, but create inconsistency when added together.

There are several common use cases for axioms in {\coq}. One is being able to use modes of reasoning that are not supported by {\coq}'s default logic: {\CompCert}%
\footnote{{\CompCert} module \modname{Axioms.v} imports module \modname{FunctionalExtensionality} from the {\coq} standard library, which both states functional extensionality and states proof irrelevance as axioms. Some {\CompCert} modules import the standard \modname{Classical} module, which states excluded-middle as an axiom. Since proof irrelevance is a consequence of excluded-middle, it should be possible to just import \modname{Classical} in \modname{Axioms.v} and deduce proof irrelevance from it.}
adds the excluded-middle ($\forall P,~P \lor \neg P$) for classical logic, functional extensionality ($f = g$ if and only if $\forall x,~f(x)=g(x)$), and proof irrelevance (one assumes that the precise statement of a proof as a $\lambda$-term is irrelevant).
Meta-theoretical arguments have shown that these three axioms do not introduce inconsistencies.%
\footnote{There is a model of {\coq}'s core calculus in Zermelo-Fraenkel set theory with the Axiom of Choice and inaccessible cardinals~\cite{DBLP:journals/corr/abs-1111-0123,timany:hal-01615123}. Such a model is compatible with these axioms.
  Previously, in times when {\coq}'s \lstinline|Set| sort was impredicative (it can still be selected to be so by a command-line option), it became apparent that this was incompatible with excluded-middle and forms of choice suitable for finding representatives of quotient sets~\cite{DBLP:conf/types/ChicliPS02,Chicli_PhD_2003}. This should be a cause of caution, though we think it unlikely to exploit such paradoxes by accident.}

Another use case for axioms is to introduce names for types, constants and functions defined in {\ocaml}, with a relationship between these and those of the {\ocaml} types and functions to be specified for {\coq}'s extraction facility.
For instance, to call an {\ocaml} function \lstset{language=[Objective]Caml}\lstinline|f: nat -> bool list| one would use
\begin{lstlisting}[language=Coq,firstline=2]
Axiom f: nat -> list bool. Extract Inlined Constant f=>"f".
\end{lstlisting}
This is used extensively in {\CompCert}, to call algorithms implemented in {\ocaml} for efficiency, using machine integers and imperative data structures; see~\ref{sec:functional}
Similarly, one can refer to an {\ocaml} constant as follows\footnote{This may allow compiling a {\coq} development once ({\coq} compilation may be expensive, certain proofs take a lot of time) and then adjust some constants when compiling and linking the extracted {\ocaml} code, maybe for different use cases.
This is not used in {\CompCert}, which, instead for flexibility, allows certain features to be selected at run-time through command-line options.
}
\begin{lstlisting}[language=Coq]
Axiom size : nat.   Extract Inlined Constant size => "size".
\end{lstlisting}

Incorrect use of axioms to be realized through extraction can lead to logical inconsistency. Consider, for instance this variant, where the \lstset{language=Coq}\lstinline|size| external definition is supposed to be a negative natural number (maybe because we mistakenly typed $n < 0$ instead of $n < 10$); one can easily derive \lstinline|False| from it:
\begin{lstlisting}[language=Coq]
Axiom size : { n : nat | n < 0 }.
\end{lstlisting}
One approach for avoiding such logical inconsistencies is to avoid axioms that specify types carrying logical specifications, that is, proofs (e.g., here $n < 0$); this is anyway a good idea, because such types may also result in mismatches (see~\ref{sec:mismatch}).
No {\ocaml} function in {\CompCert} accessed from {\coq} has {\coq} type carrying logical specification, with one exception, in {\CompCertKVX}:
\begin{lstlisting}
Axiom profiling_id: Type.
Axiom profiling_id_eq: forall (x y : profiling_id), {x=y} + {x<>y}.
\end{lstlisting}
These axioms state that there exists a type called \lstinline|profiling_id| fitted with a decidable equality, both of which are defined in {\ocaml}.
This decidable equality is a technical dependency of the decidable equality over instructions.

In order to avoid logical inconsistencies due to axioms referring to external definitions, one can prove that the type in which the \lstinline|Axiom| command states that there exists a certain term is actually inhabited; this establishes that the axiom does not introduce inconsistency. For instance, one can specify an {\ocaml} constant $n < 10$, to be resolved at compile-time, and exclude logical inconsistency by showing that such a constant actually exists:
\begin{lstlisting}[language=Coq,literate={{exists}{exists}6},firstline=3]
Axiom size : { n : nat | n < 10 }.
Lemma size_can_exist: { n : nat | n < 10 }.
Proof. exists 0; lia. Qed.
\end{lstlisting}
This approach is occasionally used in {\coq} and {\CompCert} for axiomatizing algebraic structures. For instance, {\coq} specifies constructive reals axiomatically, then provides an implementation that satisfies that specification; 
{\CompCertKVX}'s impure monad (discussed in Section~\ref{sec:functional}) is specified axiomatically, but the authors provide several implementations satisfying that specification~\cite{boulme:tel-03356701}.
Similarly, the authors could have provided an implementation of \lstinline|profiling_id|  (e.g., natural numbers) and \lstinline|profiling_id_eq| to show that these two axioms did not introduce logical inconsistencies.
    
\subsection{Mismatches between {\coq} and {\ocaml}}\label{sec:mismatch}
Though safe, the extractor can be used inappropriately.
We have just seen that adding an axiom standing for an {\ocaml} function can, if that axiom is not realizable in {\coq}, lead to logical inconsistency.
Even if the axiom is logically consistent, extraction to arbitrary {\ocaml} code can lead to undesirable runtime behavior.

An obvious case is when, in addition to an axiom specifying a constant referring, at extraction time, to an {\ocaml} function, one adds an axiom specifying the behavior of that function, and that behavior does not match the specification.
For instance, one can specify \lstinline|f| to be a function returning a natural number greater than or equal to 3, then, through extraction, define it to return~$0$:

\begin{lstlisting}
Axiom f : nat -> nat.  Axiom f_ge_3 : forall x, (f x) >= 3.
Definition g x := Nat.leb 1 (f x). 
Extract Constant f => "fun x -> O".
\end{lstlisting}

\noindent Unsurprisingly, it is possible to prove in {\coq} that $g$ always returns true, and yet to run the {\ocaml} code and see that it returns false.
It is similarly possible to write {\coq} code with impossible cases that the extractor will extract to \lstset{language={[Objective]Caml}}\lstinline|assert false|, and the extracted code will actually reach this statement and die with an uncaught exception---an after all better outcome than producing output that contradicts theorems that have been proved.
\lstset{language=Coq}
In the following code, \lstinline|False_rec _ _| eliminates on \emph{False}, which is obtained from contradiction with $x \geq 3$, and is extracted to an always failing assertion.
\begin{lstlisting}
Program Definition h x := match f x with
  | O => False_rec _ _         | S O => False_rec _ _
  | S (S O) => False_rec _ _   | S (S (S x)) => x
  end.
\end{lstlisting}

\lstset{language=Coq}
Axiomatizing the behavior of externally defined functions circumvents the idea of verified software; nowhere in the {\CompCert} source code is there such axiomatization.
An equivalent but perhaps more discreet way of axiomatizing the behavior of {\ocaml} function is through dependent types.
Consider, again,
\begin{lstlisting}
Axiom size : { n : nat | n < 10 }.
\end{lstlisting}
It is possible, through extraction mechanisms, to bind \lstinline|size| to the {\ocaml} constant $11$; this is because the type of \lstinline|size| is extracted to the same exact {\ocaml} type as \lstinline|nat|, the proof component is discarded.
It is then possible to similarly lead the {\ocaml} code extracted from {\coq} to cases that should be impossible.

The only case of such axiomatization, in {\CompCertKVX}, is the previously introduced \lstinline|profiling_id_eq| axiom, which is bound to the \lstinline|Digest.equal| function from {\ocaml}'s standard library, and defined to be string equality. We can surely assume that {\ocaml}'s string equality test to be correct, otherwise many things in {\coq} and other tools used to build {\CompCert} are likely incorrect as well.

It is also possible to instruct the extractor to extract certain {\coq} types to specific {\ocaml} types, instead of emitting a normal declaration for them.
The main use for this is to extract {\coq} types such as \lstinline|list| or \lstinline|bool| to the corresponding types in the {\ocaml} standard library, as opposed to introducing a second list type, a second Boolean type;
this is in fact so common that the standard \modname{Coq.extraction.ExtrOcamlBasic} specifies a number of such specific extractions, and so does {\CompCert}.
This is not controversial.
The extractor also allows fully specifying how a {\coq} type maps to {\ocaml}, including the constructor and ``match'' destructor;
the only use of this feature in {\CompCert} is in {\CompCertKVX} for implementing some forms of hash-consing (Sec.~\ref{sec:hash_consing}).

An in-depth discussion of further aspects of {\coq}/{\ocaml} interfacing may be found in Boulmé's habilitation thesis~\cite{boulme:tel-03356701}.

\subsection{Interfacing External Code as Pure Functions}\label{sec:functional}
{\coq} is based on a pure functional programming language; as in mathematics, if the same function gets called twice with the same arguments, it returns the same value.
{\ocaml} is an impure language, and the same function called with the same arguments may return different values over time, whether it depends on mutable state internal to the program or on external calls (user input, etc.).
By binding {\coq} axioms to impure functions, we can, again, lead {\ocaml} code extracted from {\coq} to places it should not go.

For instance, the \lstinline|z| Boolean expression extracted from this {\coq} program is \lstinline|false| though it is proved to be \lstinline|true|: it calls the same function twice with the same argument and compares the result\footnote{This result is computed by the ``\texttt{Nat.eqb}'' Boolean equality over naturals (in contrast, the {\coq} propositional equality, written ``\texttt{=}'', is only logical).}; but since that function is impure and returns the value of a counter incremented at each call, two successive calls always return unequal values.
\begin{lstlisting}[language=Coq,firstline=2,lastline=8]
Axiom f: unit -> nat.
Extract Constant f =>
  "let count = ref O in fun () -> count := S (!count); !count". 
Definition z: bool := Nat.eqb (f tt) (f tt).
Lemma ztrue: z = true.
  unfold z; rewrite Nat.eqb_refl; congruence.
Qed.
\end{lstlisting}

{\CompCert} calls a number of {\ocaml} auxiliary functions as pure functions, most notably the register allocator.
These functions are ``oracles'', in the sense that they are not trusted to return correct results; their results are used to guide compilation choices, and may be submitted to checks.
Both {\CompCertSSA} and {\CompCertKVX} add further oracles.

Could impure program constructs, in particular mutable state, in these oracles, lead to runtime inconsistencies? The code of some of these oracles is simple enough that it can be checked to behave overall functionally: mutable state, if any, is created locally within the function and does not persist across function calls. In the register allocator, there are a few global mutable variables (e.g., \lstinline|max_age|, \lstinline|max_num_eqs|), and perhaps it is possible to obtain different register allocations for the same function by running the allocator several times.
It seems unlikely that some {\CompCert} code would intentionally call a (possibly computationally expensive) oracle twice with same inputs, then go to an incorrect answer if the two returned values differ. Yet, it is not obvious that this cannot happen.

To avoid such uncertainties, the {\CompCertKVX} authors encapsulated some of their oracles, in particular oracles used within simulation checkers by symbolic execution~\cite{DBLP:journals/pacmpl/SixBM20,six:phd,six:hal-03200774},
inside the \emph{may-return monad} of~\cite{boulme:tel-03356701}. The monad models nondeterministic behavior: the same function may return different values when called with the same argument without leading into inconsistent cases.
Beyond soundness, a major feature of this approach is to provide ``theorems for free'' about polymorphic higher-order foreign {\ocaml} code. In other words, this approach ensures for free (i.e., by the {\ocaml} typechecker) that some invariants proved on the {\coq} side are preserved by untrusted {\ocaml} code~\cite{boulme:tel-03356701}. While this technique has been intensively applied within the \emph{Verified Polyhedron Library}~\cite{DBLP:conf/synasc/BoulmeMMPY18}, it is only marginally used within the current {\CompCertKVX}, only for a linear-time inclusion test between lists.

This approach however has two drawbacks.
Firstly, despite the introduction of tactics based on weakest liberal precondition calculus, the proof effort is heavier than for code written with pure functions without a monadic style.
Secondly, all the code calling impure functions modeled within the may-return monad also becomes impure code modeled within that monad, meaning that a significant part of the rest of {\CompCert} (at least the code calling the sequence of optimization phases and their proofs) would have to be rewritten using that monad.%
\footnote{Much of {\CompCert} is already written in an error monad, with respect to which, the may-return monad is a straightforward generalization. It thus seems feasible to rewrite {\CompCert} with the may-return monad instead of the existing error monad. In practice, this represents a lot of reengineering work. For example, currently, the may-return monad provides a tactic in backward reasoning, based a weakest-precondition calculus. In contrast, {\CompCert} provides a tactic for forward reasoning on the error monad. Thus, defining a tactic on the may-return monad that behaves like the one of the error monad would help in reducing the amount of changes in {\CompCert} proofs.}

{\CompCert}'s {\coq} code accesses mutable variables storing command-line options through helper functions. This supposes that these variables stay constant once the command line has been parsed, which is the case.

\lstset{language=Coq}
In {\coq}, all functions must be shown to be terminating (because nonterminating terms can be used to establish inconsistencies). Arguments for the termination of a function are sometimes more intricate and painful to write in {\coq} than those for its partial correctness, and termination is not really useful in practice: from the point of view of the end-user there is no difference between a terminating function that takes prohibitively long time to terminate, and a nonterminating function.
For this reason, some procedures in {\CompCert} and forks that search for a solution to a problem (e.g., a fixpoint of an operator) are defined by induction on a positive number, and return a default or error value if the base case of the induction is reached before the solution is found.
\modname{Iteration.PrimIter}, used for instance in the implementation of Kildall's fixpoint solving algorithm for dataflow analysis, thus uses a large positive constant \lstinline|num_iterations|=$10^{12}$.
Such numbers are often informally known as~\emph{fuel}.

{\CompCertSSA} takes an even more radical view: a natural number \lstinline|fuel| is left undefined, as an axiom, inside the {\coq} source code, and is extracted to {\ocaml} code
\lstset{language=[Objective]Caml}\lstinline|let rec fuel = S fuel|, meaning that \lstinline|fuel| is circularly defined as its own successor, and in practice acts as an infinite stream of successors.
Why that choice?
\lstset{language=Coq}
\lstinline|num_iterations| is a huge constant belonging to the \lstinline|positive| type, which models positive integers in binary notation; there is a custom induction scheme for this type that implements the usual well-founded ordering on positive integers.
In contrast, \lstinline|fuel| is a natural number in unary notation, on which inductive functions may be defined by structural induction, which is a bit easier than with a custom induction scheme;
but it is impossible to define a huge constant in unary notation.
The \lstinline|num_iterations| scheme is cleaner, but we have not identified any actual problem with the \lstinline|fuel| scheme.
The {\ocaml} code extracted from {\coq} has no way to distinguish \lstinline|fuel| from a large constant.

The \lstinline|fuel| trick however breaks if pointer equality is exposed on the natural number type~\cite{boulme:tel-03356701}.
The following program, defined using a ``may return'' monad, where \lstinline|phys_eq_nat| is pointer equality on natural numbers, can be proved not to return true; yet, it does return true at runtime.
\begin{lstlisting}
Definition fuel_eq_pred :=
  match fuel with
  | O => Impure.ret false
  | S x => phys_eq_nat fuel x
  end.
\end{lstlisting}

\subsection{Pointer Equality and Hash-Consing}\label{sec:hash_consing}
The normal way in {\coq} to decide the equality of two tree-like data structures is to traverse them recursively.
The worst-case of this approach is reached when the structures are equal, in which case they will be traversed completely.
Unfortunately this case is frequent in many applications for verified compilation, verified static analysis, etc.:
when the data structures represent abstract sets of states (in abstract interpretation), equality signals the equality of these abstract sets, which indicates that a fixed point is reached;
equality between symbolic expressions is used for translation validation through symbolic execution~\cite{DBLP:journals/pacmpl/SixBM20}.
Furthermore, there are many algorithms that traverse pairs of tree-like structures for which there are shortcuts if two substructures are equal: for instance, if this algorithm computes the union of two sets, then if these sets are equal, then the union is the same~\cite[\S5]{DBLP:conf/lctrts/MonniauxS21};
being able to exploit such cases has long been known to be important for the speed of static analyzers~\cite[\S6.1.2]{BlanchetCousotEtAl_PLDI03}.

\lstset{language={[Objective]Caml}}
If we were programming in {\ocaml}, we could simply use pointer equality (\lstinline|==|) for a quick check that two objects are equal: if they are at the same memory location, then they are necessarily structurally equal (the converse is not true in general).
In {\coq}, a naive formalization of this approach could be:
\begin{lstlisting}[language=Coq]
Parameter A: Type.
Axiom phys_eq: A -> A -> bool.
Axiom phys_eq_implies_eq: forall x y, phys_eq x y = true -> x = y.
\end{lstlisting}

\lstset{language=Coq}
This approach is however unsound.%
\footnote{We saw in the preceding section another possible cause of unsoundness: if circular data structures are defined in {\ocaml} inside inductive types, pointer equality can be used to establish that a term is equal to one of its strict subterms, which is normally impossible, thus leads to an absurd case at execution time. To avoid this, either completely disallow linking to circular terms constructed in {\ocaml}, or restrict pointer equality test to types where such circular terms are not constructed.}
We prove that \lstinline|x_eq_x| and \lstinline|x_eq_y| are equal; yet in the extracted code, the former evaluates to true, the second to false.
\begin{lstlisting}[language=Coq,firstline=5]
Definition x     :=S O. (* 1 *) Definition y     :=S O. (* 1 *)
Definition x_eq_x:=phys_eq x x. Definition x_eq_y:=phys_eq x y.

Extract Inlined Constant phys_eq => "(==)".
Recursive Extraction x_eq_x x_eq_y.
Lemma same : x_eq_x = x_eq_y.         Proof. reflexivity. Qed.
\end{lstlisting}
\lstset{mathescape=true}
To summarize, {\ocaml} pointer equality can distinguish two structurally equal objects, whereas this is provably impossible for {\coq} functions: 
for {\coq}, \lstinline|x| and \lstinline|y| are the same, so they are interchangeable as arguments to \lstinline|phys_eq|.
This is the functionality issue of Section~\ref{sec:functional} in another guise: the same {\ocaml} function must be allowed to return different values when called with the same argument.

The solution used in {\CompCertKVX} for checking that symbolic values are equal was thus to model pointer equality as a nondeterministic function in a ``may return'' monad. In this model~\cite{boulme:tel-03356701},
pointer equality nondeterministically discovers some structural equalities.\footnote{In this model, a given {\coq} term is not necessarily equal to ``itself'' for pointer equality, because, in a {\coq} proposition, ``itself'' implicitly means a structural copy of ``itself''.}
This solution has one drawback: the whole of the symbolic execution checker is defined within this monad, and the authors unsafely exit from that monad to avoid running much of {\CompCert} through it.
It is uncontroversial that pointer equality implies equality of the pointed objects. The only cause for unsoundness in such an approach could be the unsafe exit. Yet, again, why would {\CompCertKVX} call twice the symbolic execution engine with the same arguments to reach an absurd case for different outcomes?

Opportunistic detection of identical substructures through pointer equality was implemented for instance in Astrée~\cite{BlanchetCousotEtAl_PLDI03}.
This approach takes advantage of the fact that many algorithms operating on functional data structures simply copy pointers to parts of structures that are left intact:
The opportunistic approach detects that some parts of structures have been left untouched, skipping costly traversals.
It however does not work if a structure is reconstructed from scratch, for instance as the result of a symbolic execution algorithms:
if two symbolic executions yield the same result, these results are defined by isomorphic data structures but the pointers are different.
What is needed then is \emph{hash-consing}: when constructing a new node, search a hash-table containing all currently existing nodes for an identical node and return it if it exists, otherwise create a new node and insert it into the table.
Hash-consing is widely used in symbolic computation, SMT-solvers etc.; there exist libraries making it easy in {\ocaml}~\cite{DBLP:conf/ml/FilliatreC06}, and the {\ocaml} standard library contains a weak hash-table module, one of the main uses of which is being a basic block for hash-consing.

The difficulty is that, though overall the construction of new objects behaves functionally (it returns objects that are structurally identical to what a direct application of a constructor would produce), it internally keeps a global state inside the hash-table.
Several solutions have been proposed to that problem~\cite{Braibant_Jourdan_Monniaux_JAR2014};
one is to keep that global state explicitly inside a state monad, which amounts to threading the current state of the hash table through all computations.
In the original version from~\cite{Braibant_Jourdan_Monniaux_JAR2014}, this implied implementing the hash-table by emulating an array using functional data structures, which was very inefficient.
{\coq}~8.13 introduced primitive 63-bit integers and arrays (with a functional interface), optimized for cases where the old version of an updated array is never used anymore~\cite[\S2.3]{DBLP:conf/ml/ConchonF07}, which, through special extraction directives, may be extracted to {\ocaml} native integers and arrays.
That solution was not adopted for {\CompCertKVX}, only because {\coq}~8.13 had not yet been released when the project started.
Instead, {\CompCertKVX} has experimented with two alternative approaches for hash-consing.

The first approach used in {\CompCertKVX} introduces an untrusted {\ocaml} function (modeled as a nondeterministic function within the may-return monad) that constructs terms through the hash-consing mechanism (searching in the hash-table etc.); these terms are then quickly checked for equivalence with the desired terms, using a provably correct checker.
For instance, if a term $c(a_1,\dots,a_n)$ is to be constructed, and the function returns a term $t$, then the root constructor of $t$ is checked to be $c$, then the arguments to that constructor are checked to be equal to $a_1,\dots,a_n$ by pointer equality.%
\footnote{A unique identifier is added as an extra field to each object, for reasons including efficient hashing. Structural equality is thus modulo differences in unique identifiers.}  
This solution does not add anything to the trusted computing base, apart from pointer equality.
A may-return monad is used because the {\ocaml} code is untrusted, and in particular is not trusted to behave functionally.
The drawback is that, though the {\ocaml} code will always make sure that there are never two identical terms in memory at different pointer addresses, this is not reflected from the point of view of proofs: in the
{\coq} model (discussed above) of pointer equality within the may-return monad, pointer equality implies structural equality, but structural equality does not imply pointer equality.
However, only the former is needed for a symbolic execution engine that checks that two executions are indeed equivalent by structural equality of terms, as in the scheduler in {\CompCertKVX}~\cite{DBLP:journals/pacmpl/SixBM20}.

Having to thread a whole computation through a monad, further adding to proof complexity, for actions that are expected to behave functionally overall, is onerous.
One solution is to add hash-consing natively inside the runtime system;
for instance, the \software{GimML} language,%
\footnote{\url{https://projects.lsv.fr/agreg/?page_id=258} Formerly HimML.}
from the ML family~\cite{GimML_refman,Goubault94himml:standard,JG:Sharing}, automatically performs hash-consing on datatypes on which it is safe to do so, which is for instance used to implement efficient finite sets and maps.
This can be emulated by a ``smart constructor'' approach~\cite{Braibant_Jourdan_Monniaux_JAR2014}, replacing, through the extraction mechanism, calls to the term constructor, term pattern matching, and term equality by calls to appropriate {\ocaml} procedures: the constructor performs hash-consing, the pattern matcher performs pattern matching ignoring the internal-use ``unique identifier'' field used for hash-consing, and term equality is defined to be pointer equality;
appropriate {\ocaml} encapsulation prevents manipulation of these terms except through these three functions, and in particular prevent them from being constructed by other methods than the smart constructor.   
Assuming that this {\ocaml} code is correct, this is indeed sound, due to the global invariant that there never exist two distinct yet structurally identical terms of the hash-consed type currently reachable inside memory.
Because terms can only be built using the smart constructor, and that hash-consing ensures that pointer equality is equivalent to structural equality, pointer equality can indeed be treated as a deterministic function, without need for a monad.
This approach has the benefit of an easy-to-understand interface and simple proofs;
this was the second approach experimented within {\CompCertKVX} and was used for the \modname{HashedSet} module~\cite{DBLP:conf/lctrts/MonniauxS21}.

This second approach adds significantly more {\ocaml} code to the trusted computing base than just assuming that pointer equality implies structural equality.
Yet, this {\ocaml} code is small, with few execution paths, and can be easily tested and audited. It assumes the correctness of {\ocaml}'s weak hash-tables; however, {\coq}'s kernel includes a module (\modname{Hashset}) that is also implemented using these weak hash-tables, so one already assumes that correctness when using~{\coq}.

\section{Front-end and semantic issues}\label{sec:frontend}
{\CompCert} parses C and assigns a formal semantics to it. As such, it depends on a formal model of the C syntax and a formal semantics for it, supposed to reflect the English specification given in the international standard~\cite{C11}. {\CompCert} supports an extensive subset of C99~\cite{C99} (notable missing items are variable-length arrays and some forms of unstructured branching, à la Duff's device) and some C11 features (note that in C11, support for variable-length arrays is optional).%
\footnote{The {CH\textsubscript{2}O} project (\url{https://robbertkrebbers.nl/research/ch2o/}) aims at formalizing the ISO C11 standard in {\coq}. This development is unrelated to the formalization inside {\CompCert}.}

The formal semantics of C supported by {\CompCert} is called ``{\CompCert}~C''. Converting the source program, given in a text file, to the {\CompCert}~C AST (abstract syntax tree) on which the formal semantics is defined, relies on many nontrivial transformations: preprocessing, lexing (lexical analysis), parsing (AST building) and typechecking.
Most of them are unverified, but trusted. There are two important exceptions: significant parts of the parser and the typechecker of {\CompCert}~C are formally verified. The formally verified parser is implemented using the \progname{Menhir} parser generator, and there is a formal verification of its correctness with respect to an attribute LR(1) grammar~\cite{Jourdan-Pottier-Leroy}. It relies on an unverified ``pre-parser'' to distinguish identifier types introduced by \lstinline|typedef| from other identifiers
(a well-known issue of context-free parsing of C programs).
It produces an AST which is then simplified and annotated with types, by another unverified pass, called ``\emph{elaboration}''.
Finally, the resulting {\CompCert}~C program is typechecked, by the formally verified typechecker. This is where the fully verified frontend of {\CompCert} really starts.

Obviously, a divergence between the semantics of C as understood by {\CompCert} and that semantics as commonly understood by programmers to be compiled may lead to problems.
Validating such semantics is an important issue~\cite{blazy:inria-00292043}.
The standard has evolved over time for taking into account common programming practices or for solving some contradictions.%
\footnote{See an example on {\scriptsize \url{http://www.open-std.org/jtc1/sc22/wg14/www/docs/dr_260.htm}}.}
{\CompCert} semantics has also evolved to get closer to the standard, see~\cite{Krebbers-Leroy-Wiedijk-2014}. 
In the last years, a few minor divergences have been spotted.
For instance, there was a minor misimplementation of scoping rules (commit~\compcertcommit{99918e4}{99918e4118e0ea644b20e37a13ceb31d935fdda5}) that led the following program to allocate \lstinline|s| of size 3 (\lstinline|sizeof(t)| being interpreted with \lstinline|t| the global variable, whereas the standard mandates it should refer to the \lstinline|t| variable declared before it on the same line) instead of 4:
\begin{lstlisting}[language=C]
char t[]={1,2,3};
int main() { char t[]={1,2,3,4}, s[sizeof(t)];
  return sizeof(s); }
\end{lstlisting}
Another example: {\CompCert} and other compilers accepted some extension to the syntax of C99 (anonymous fields in structures and unions) but assigned slightly different meanings to it (different behavior during initialization, issue~\compcertissue{411}).

The C standard leaves many behaviors \emph{undefined}---anything can happen if the program exercises such a behavior (the compiler may refuse the program, the program may compile and run but halt abruptly when encountering the message, or may continue running with arbitrary behavior).
Some undefined behaviors, such as array access out of bounds, are exploited in malicious attacks.
The C standard also leaves many behaviors \emph{unspecified}, meaning the compiler may choose to implement them arbitrarily within a certain range of possibilities---e.g., the order of evaluation of parts of certain expressions with respect to side effects.%
\footnote{This should not be confused with syntactic associativity, which is fully defined by the standard.}
Actually, distinguishing between \emph{unspecified} and \emph{undefined} behavior in the evaluation order is rather complex: see~\cite{DBLP:journals/jar/Krebbers16} for a formal semantics.
Furthermore, many compilers implement extensions to the standard. Some deviate from the standard's mandated behavior in some respects.%
\footnote{For instance, Intel's compiler, at least at some point, deliberately deviated from standard floating-point behavior to produce more efficient code. An option was needed to get standard compliance. In contrast, {\gcc} would by default comply with the standard, and enable optimizations similar to Intel's when passed options such as \texttt{-ffast-math} or the aptly-named \texttt{-funsafe-math-optimizations}~\cite{Monniaux_TOPLAS08}.}

Many programs, be them applications, libraries or system libraries, rely on the behavior of the default compiler on their platform (e.g., {\gcc} on Linux, {\clang} on MacOS, Microsoft Visual Studio for Windows).%
\footnote{On Linux, compiling software with \texttt{gcc -std=c99}, which disables some GNU-specific extensions, often fails. On the KVX, {\CompCertKVX} includes a kludge for defining a \lstinline|__int128| type suitable enough for processing system header files.}
If compilation just fails, then issues are relatively easy (though maintaining support for multiple compilers, often through conditional compilation and preprocessor definitions, is error-prone); subtler problems may be encountered when software compiles but has different behavior with different compilers.%
\footnote{As an example, C compilers are allowed to replace \lstinline|a*b+c| by a fused multiply-add \lstinline|fma(a, b, c)|, which may produce slightly different results. Such replacements may be disabled by a command-line option or a pragma.}
It may be difficult to narrow differences in outcomes to a bug (including reliance on undefined behavior) or to a difference in valid implementations of unspecified behavior.

The only semantic issue that we know of regarding {\CompCert}'s forthcoming version 3.10 is with respect to bitfields. A write to a bitfield is implemented using bitshift and bitwise Boolean operations, and these operations produced the ``undefined'' value if one of their operands is ``undefined''. Writing to a bitfield originally stored in an uninitialized machine word or long word, which is the case for local variables, thus results in an ``undefined'' value, whereas the bits written to are actually defined. Reading from that bitfield will then produce the ``undefined'' value, as can be witnessed by running the program in {\CompCert}'s reference interpreter, which stops complaining of undefined behavior.
Fixing this issue would entail using a bit-wise memory model (issue \compcertissue{418}).%
\footnote{Questions of ``undefined'' and ``poison'' values are notoriously difficult to get right in semantics; see~\cite{DBLP:conf/pldi/LeeKSHDMRL17} for a discussion of intricate bugs in LLVM.}
It may be possible to write and prove correct a phase that would replace this ``undefined'' value by an arbitrary value and thus result in miscompilation.
We do not know, however, of any phase that would produce this in {\CompCert} or variants.

{\CompCertKVX}'s test suite includes calling compiler fuzzers CSmith%
\footnote{\url{https://github.com/csmith-project/csmith} and~\cite{YangCER11}}
and YarpGen:%
\footnote{\url{https://github.com/intel/yarpgen}}
random programs are generated, compiled with {\gcc} and {\CompCertKVX} and run on a simulated target---an error is flagged if final checksums diverge.

Due to possible semantic differences for the subset of the C language between the tools that they use for their formal proofs and {\CompCert}, Gernot Heiser, lead designer of the \software{seL4} verified kernel, \href{https://sel4.discourse.group/t/compiling-sel4-with-compcert/115}{argues} that translation validation of the results of black-box compilation by {\gcc} is a safer route:
\begin{quote}
  [\ldots] using CompCert would not give us a complete proof chain. It uses a different logic to our Isabelle proofs, and we cannot be certain that its assumptions on C semantics are the same as of our Isabelle proofs.
\end{quote}

Another option, for C code produced from a higher-level language by code generators, is to replace {\CompCert}'s frontend by a verified a code generator for that language, directly targeting one of {\CompCert}'s intermediate representations (e.g., \modname{Clight}) and semantics, as done for instance for Velus~\cite{Bourke-BDLPR-2017} for a subset of the Lustre synchronous programming language.

Some features of the C programming language are not supported by {\CompCert}'s formally verified core, but can be supported through optional unverified preprocessing, chosen by common line options:
\texttt{-fstruct-passing} allows passing structures (and unions) as value as parameters to functions, as well as returning them from a function;%
\footnote{In C, passing \emph{pointers} to structures that container parameters or are meant to container return values is a common idiom. The language however also allows passing or returning the structures themselves, and this is implement in various ways by compilers, including passing pointers to temporary structures or, for structures small enough to fit within a (long) machine word, directly as an integer register.
  How to do so on a given platform is specified by the ABI.\@%
Parameter passing, with all particular cases, may be a quite delicate and convoluted part of the ABI.}
\texttt{-fbitfields} allows bit fields in structures.%
\footnote{Recently, direct verified handling of bitfields was added to {\CompCert} (commit~\compcertcommit{d2595e3}{d2595e3afb8c38a3391a66c3fc3f7a92fff9eff4}). This should be available in release~3.10.}
Preprocessing implements these operations using lower-level constructs (memory copy builtin, bit shift operators), sometimes in ways incompatible with other compilers---{\CompCert}'s manual details such incompatibilities.

In addition, option \texttt{-finline-asm} allows inline assembly code with parameter passing, in a way compatible with {\gcc} (implementing a subset of {\gcc}'s parameter specification). The semantics of inline assembly code is defined as clobbering registers and memory as specified, and emitting an externally observable event.
Option \texttt{-fall} activates structure passing, bitfields, and inline assembly, for maximal compatibility with other compilers.

Because inline assembly is difficult to use,%
\footnote{Inline assembly is so error-prone that specialized tools have been designed to check that pieces of assembly code match their read/write/clobber specification~\cite{DBLP:conf/icse/RecoulesBBLMP21}.}
and because its semantics involves emitting an event, preventing many optimizations, {\CompCert} also provides builtin functions that call specific processor instructions. If a builtin has been given an arithmetic semantics, then it can be compiled into arithmetic operators suitable for optimization;
this is the case, for instance, of the ``fused multiply add'' operator on the KVX.\@%
In contrast, instructions that change special processor registers are defined to emit observable events.

\section{Assembly back-end issues}\label{sec:assembly}
The verified parts of {\CompCert} do not output machine code, let alone textual assembly code. Instead, they construct a data structure describing a set of global definitions: variables and functions; a function contains a sequence of instructions and labels.
The instructions at that level may be actual processor instructions, or pseudo-instructions, which are expanded by unverified {\ocaml} into a sequence of actual processor instructions.
The resulting program is printed to textual assembly code by the \lstinline|TargetPrinter| module;
most of it consists in printing the appropriate assembly mnemonic for each instruction, together with calling functions for printing addressing modes and register names correctly, but there is some arcane code dealing with proper loading of pointers to global symbols, printing of constant pools, etc.
Some of this code depends on linking peculiarities and on the target operating system, not only on the target processor.

\subsection{Printing Issues}\label{sec:prettyprint}
An obvious source of potential problems is the huge ``match'' statement with one case per instruction, each mapping to a ``print'' statement. If the ``print'' statement is incorrect, then the instruction printed will not correspond to the one in the data structure. Printing an ill-formed instruction is not a serious problem, as the assembler will refuse it and compilation will fail. There have however been recent cases where {\CompCert} printed well-formed text assembly instructions that did not correspond to the instruction in the data structure. The reason why such bugs were not caught earlier is that these instructions are rarely used.
Commit~\compcertcommit{2ce5e496}{2ce5e496b8d4c838c87c9f00a84ed23d1abc26fc} fixed a bug resulting in some fused multiply-add instructions being printed with arguments in the wrong order; these instructions are selected only if the source code contains an explicit fused multiply-add builtin call, which is rare.
In {\CompCertKVX}, commit~\compcertkvxcommit{e2618b31}{e2618b31dac9aa0cd859466b0e6af13ed00dc877} fixed a bug---``nand'' instructions would be printed as ``and''; ``nand'' is selected only for the rare \lstset{language=C}\lstinline|~(a & b)| pattern. The bug was found by compiling randomly generated programs.

In some early versions of {\CompCert} there used to be a code generation bug~\cite[\S3.1]{YangCER11} that resulted in an exceedingly large offset being used in relative addressing on the PowerPC architecture; this offset was rejected by the assembler.
Similar issues surfaced later in CakeML on the MIPS-64 architecture~\cite{DBLP:conf/cpp/FoxMTK17} and in {\CompCert} on AArch64 (commit \compcertcommit{c8ccecc}{c8ccecc783671fb699a33f432c34e3c1cd1dc801}).
This is a sign that constraints on immediate operand sizes are easily forgotten or mishandled,%
\footnote{For instance, {\CompCertKVX} generates loads and stores of register pairs on AArch64, with special care: their offset range is smaller than for ordinary loads and stores.} 
and a caution: incorrect value sizes could result in situations not resulting in assembler errors.

\subsection{Pseudo-Instructions}\label{sec:pseudoinst}
In addition to instructions corresponding to actual assembly instructions, the assembler abstract syntax in {\CompCert} features pseudo-instructions, or macro-instructions, most notably:
allocation and deallocation of a stack frame;
copying a memory block of a statically known size;
jumping through a table.
The reasons why these are expanded in unverified {\ocaml} code are twofold.
First, the correspondence between the semantics of such operations and their decomposition cannot be easily expressed within {\CompCert}'s framework for assembly-level small-step semantics, especially the memory model.
{\CompCert} models memory as a set of distinct blocks, and pointers as pairs (block identifier, offset within the block);
\footnote{This reflects the C standard's view that variables and blocks live each in their own separate memory space.
  For instance, in C, comparisons between pointers to distinct variables have undefined behavior~\cite[\S6.5.8]{C11}.
Some {\CompCert} versions in which pointers truly are considered to be integers have been proposed~\cite{DBLP:journals/jar/BessonBW19a,10.1145/2980983.2908109}.}
stack allocation and deallocation create or remove memory blocks by moving the stack pointer, which is just a positive integer.
Jump tables (used for compiling certain \lstinline|switch| statements) are arrays of pointers to instructions within the current function, whereas {\CompCert} only knows about function pointers.
Second, their expansion may use special instructions (load/store of multiple registers, hardware loops\dots) not normally selected, the behavior of which may be difficult to express in the semantics%
\footnote{Hardware loops, on processors such as the KVX, involve special registers. When the program counter equals the ``loop exit'' register, and there remain loop iterations to be done, control is transferred to the location specified by the ``loop start'' register.
  In all existant {\CompCert} assembly language semantics, non-branching instructions go to the next instruction.
  Modeling hardware loops would thus involve changing all instruction semantics to transfer control according to whether the loop exit is reached, proving invariants regarding the hardware loop registers, etc.
  This could be worth it if the hardware loops could be selected for regular code, not just builtins, but this itself would entail considerable changes in
  previous compiler phases.}
or the memory model. This is typically the case for memory copy; see below.

\paragraph{Stack Frame (De)Allocation}
Stack (de)allocation pseudo-instructions address the  gap between the abstract representation of the memory as a set of blocks completely separated from each other
and the flat addressing space implemented by most processors, call frames laid out consecutively, allocation and deallocation amounting to subtracting or adding to the stack pointer.
A refined view, with a correctness proof going to the flat addressing level, was proposed for the x86 target~\cite{10.1145/3290375} but not merged into mainline~{\CompCert}.%

\paragraph{Loading Constants}
Certain instructions may need some expansion and case analysis, and possibly auxiliary tables. For instance, on the ARM architecture, long constants must be loaded from constant pools addressed relatively to the program counter; thus emitting a constant load instruction entails emitting a load and populating the constant pool, which must be flushed regularly since the range of adressing offsets is small.
Getting the address of a global or local symbol (global or \lstinline|static|) variable may also entail multiple instructions, and perhaps a case analysis depending on whether the code is to be position-independent, and, in {\CompCertKVX}, whether the symbol resides in a thread-local program section.%
\footnote{In C11~\cite{C11}, the \lstinline|_Thread_local| storage class specifies that one separate copy of the variable exists for each thread.
Typically, a processor register points to the thread-local memory area and these variables are accessed by offsets from that register.
{\CompCert} has no notion of concurrency, but on the KVX, some system variables are thread-local and must be accessed as such even from single-threaded programs.}
The low-level workings of the implementation of these pseudo-instructions rely on the linker performing relocations, on the application binary interface specifying that certain registers point to certain memory sections, etc.

\paragraph{Builtins}
{\CompCert} allows the user to call special ``builtins'', dealing mainly with special machine registers and instructions (memory barriers, etc.). These builtins are expanded in \modname{Asmexpand} or \modname{TargetPrinter} into actual assembly instructions.

As an example, consider the memory copy builtin, which may both be used by the user (with \lstinline|_builtin_memcpy_aligned()|) to request copying a memory block of known size, and is also issued by the compiler for copying structures.
Expanding that builtin may go through a case analysis on block size and alignment: smaller blocks will be copied by a sequence of loads and stores, larger blocks using a loop. The scratch registers may be different in each case, and this case analysis must be replicated in the specification; alternatively, the specification may contain a upper-bound on the set of clobbered registers, but in any case no clobbered register should be forgotten.
There may also be a complicated distinction of cases regarding which source register is alias to which other source register, or which scratch one.
A bug in that builtin, which did not check alignment and generated improper offsets for load instructions, was found in {\CompCert} on AArch64; the assembler would reject the generated code (commit \compcertcommit{c8ccecc}{c8ccecc783671fb699a33f432c34e3c1cd1dc801}).
Another bug in the same builtin, on four architectures (ARM, AArch64, PowerPC, RISC-V), due to an incorrect test about register aliasing, resulted in successful compilation, assembly and linking with incorrect code being emitted (commit \compcertcommit{c2c871c}{c2c871c78d4021a28be8ba5c2d8454cfc10fad22}).

One bug was found in the {\CompCertKVX} stack frame allocation code, which had no adverse consequence unless a very large stack frame or many parameters were used, which explains why it was not detected earlier~(commit \compcertkvxcommit{fccfa9}{fccfa9b6ac74953af188a2538eb9cd7258544c1a}).

\paragraph{Clobbered Registers}
Expansions of pseudo-instructions and builtins often use scratch registers. The registers that are clobbered by each pseudo-instruction and builtin are defined in the {\coq} file (\modname{Asm.v}) giving the semantics of the abstract assembly language. Thus, changes to expansions must affect coherently both the \modname{Asm.v} specification and the \modname{AsmExpand} and/or \modname{TargetPrinter} {\ocaml} module.

In the last few years, several specification bugs about registers clobbered by pseudo-instructions and builtins were found in {\CompCert}, on several architectures. Commit~\compcertcommit{0df99dc4}{0df99dc46209a9fe5026b83227ef73280f0dab70} fixes several wrong specifications of clobbered registers on AArch64; commit~\compcertcommit{a4cfb9c2}{a4cfb9c2ffdef07fa0d478e66f279687c9823d42} on ARM;\@commit~\compcertcommit{39710f78}{39710f78062a4a999c079b58181a58e62b78c30b} on RISC-V.
It seems that none of these bugs could result in the generation of incorrect code, for the registers that were wrongly specified not to be clobbered were not used by the {\CompCert} code generator to store persistent data.
The problem is that it was possible to modify the code generator with full correctness proof, and have {\CompCert} generate incorrect code.
For instance, some pseudo-instructions would use the return address register as a scratch register, not specified as clobbered.
Some compilers perform leaf function optimization: the prologues and epilogues of functions that never call other functions do not save and restore the return address.
{\CompCert} applies this optimization only on the PowerPC architecture, and even then only partially;
if one had added this optimization to AArch64 or RISC-V, incorrect code would be generated in leaf functions using the wrongly specified pseudo-instructions, though all proofs would go through.

Bugs in expansion of builtins due to incorrect specification of clobbered registers (or memory), and those related to outcome depending on compiler choices (e.g., register aliases), eerily resemble those due to improper use of inline assembly in C programs~\cite{DBLP:conf/icse/RecoulesBBLMP21}. Perhaps similar methods of validation could be used.

As an alternative, we propose moving the parts that deal with case distinctions (register aliasing, sizes, alignments\dots) out of the untrusted code base into the trusted code base, possibly one pseudo-assembly instruction for each case.
For instance, there could be one ``memory copy'' pseudo-assembly instruction for each different code sequence to be generated, with fixed ``clobbered'' registers and explicit constraints on alignment, size etc.\ in the specification of the instruction.
Verified {\coq} code would select the proper pseudo-instruction to use.
This would likely avoid bugs due to case distinctions in trusted code, alleviate difficulties in properly specifying the pseudo-instructions and keeping this specification synchronized with their expansion, and make it easier to perform unit testing on the expansions.

\subsection{Microarchitectural Concerns}
{\CompCertKVX} adds two levels of instruction scheduling to {\CompCert} (prior and after register allocation).%
\footnote{\label{foot:fixoct2022}Tristan \& Leroy~\cite{2008-Tristan-Leroy-POPL} had previously developed scheduling for {\CompCert} but their developments were not integrated into {\CompCert} releases.
  In the proceedings version of this paper, we erroneously reported that these developments were not publicly available;
  in fact they were made available from
  \url{https://github.com/jtristan/CompCert-Extensions}.\\
  {\CompCertKVX} uses hash-consing to solve certain efficiency issues present in that early development, at the expense of extra correctness issues (see Sec.~\ref{sec:hash_consing}).}
Instruction scheduling reorders instructions while preserving semantics so as to minimize execution time.
Current high-performance processors dynamically reorder instructions, but this is complex and consumes extra energy; \emph{in-order} processors need the compiler to schedule instructions for good performance, taking into account latencies (the number of clock cycles between the operands of an instruction being read and the results being produced) and resource constraints (the number of instructions that can be simultaneously executed; e.g., a processor may be able to execute two instructions at a time, but only one of them may be a memory access, and only one of them may be floating-point).

Tables of resource uses and latencies are cumbersome to build, and often involve access to private documentation and/or reverse engineering; there are thus likely incorrect.%
\footnote{The {\CompCertKVX} team had private documentation on the KVX;\ despite that, due to the tedium of building tables, they had a few bugs, as shown by commit logs. Their tables for AArch64 and RISC-V are based on the source code of other compilers.}
Fortunately, all targets of {\CompCertKVX} have \emph{interlocked} pipelines, meaning that, if a value is read from a register that awaits a write, the instruction is stalled; thus sequential semantics are preserved: the worst that can happen if incorrect latencies are used is that the pipeline stalls for some cycles, which is a performance, not a correctness, issue. In contrast, on processors with non-interlocked pipelines the latencies belong to the semantic definition of the assembly code: a read from a register that awaits a write yields the previous value held in that register.
Regarding resource constraints, on a very large instruction word (VLIW) processor, bundles of instructions that exceed resource constraints will be refused by the assembler;
on a conventional multiple-issue processor, successive instructions that cannot be issued at the same cycle for lack of resources will be issued sequentially, which is equivalent since the processor preserves sequential semantics even when issuing several instructions.
We conclude that pipeline modeling issues have no impact on the correctness of the generated code of {\CompCertKVX}, but solely on its performance.%
\footnote{The situation would of course be very different in the case of a tool bounding worst case execution time through precise processor modeling.}

\subsection{Assembling and Linking}
{\CompCert} produces assembly code in textual form, which must then be assembled and linked using another toolchain, such as \texttt{gcc} (the GNU Compiler Collection) or \texttt{clang} (LLVM). This toolchain is thus within the TCB.
Absint GmbH, which sells the commercial releases of {\CompCert}, also sells for certain architectures the \software{Valex} tool which matches the {\CompCert} code to the binary code~\cite{Leroy-BKSPF-2016,Kastner-LBSSF-2017}.
An alternative is direct generation of machine code, as in CakeML~\cite{DBLP:conf/itp/KumarMTM18}; {\CompCertELF} extends {\CompCert} with a verified assembler for the x86 target~\cite{10.1145/3428265}.

Finally, {\CompCert}'s correctness proof was originally meant for a ``closed world'': a program wholly compiled with it as a single module.
In reality, most large C projects are compiled from multiple files which are then linked.
The correctness proof was later extended, in version~2.7, to account for separate compilation and linking, following~\cite{10.1145/2914770.2837642}.
There have been proposals for more ambitious formalizations of the linking process~\cite{10.1145/3371091}, even implementing a verified linker for a subset of ELF on the x86-32 architecture~\cite{10.1145/3428265};
\footnote{ELF is a standard file format for object code.}
Specifying and proving correct a general ELF linker is itself a fairly ambitious project~\cite{DBLP:conf/oopsla/KellMS16}.

\section{Modeling and Application Binary Interface Issues}\label{sec:ABI}
The semantics of assembly instructions is defined, for each architecture, in the official manuals from the architecture designers.
The \emph{application binary interface} (ABI), specific to each combination of architecture and operating system (or execution environment), defines how parameters are to be passed (in which registers, etc.), what kind of different global symbols exist and how they are accessed, what registers are reserved for system use, how the execution stack is to be laid out, what values the high-order bits of long registers may contain if the register contains a shorter value, etc.
In contrast, {\CompCert}'s vision of values is somewhat abstract, even at the assembly level, which may pose problems especially when interfacing to other parts of the runtime system.

\subsection{Modeling of Values}
{\CompCert} considers that a value, e.g., stored in a register, is either a 32-bit integer; a 64-bit integer; a 32-bit single precision floating-point number; a 64-bit double precision floating-point number; a pointer, consisting in a block identifier and an offset; or ``undefined'', a value that can be refined into any other value, modeling undefined behavior that does not stop program execution (because not yet externally observed).
This is, however, an abstraction of reality.
Pointers, in reality, are not a pair (block, offset) but a single 32-bit or 64-bit integer.
How is a 32-bit value stored in a 64-bit register? Are the higher-order bits indifferent, supposed to be $0$ ($0$-extension) or equal to the sign bit (sign-extension)?

These modeling issues have subtle consequences on the implementation of certain instructions.
If the application binary interface specifies that 32-bit values stored in 64-bit processor registers are $0$-extended, then the $0$-extension operation as defined in {\CompCert} (taking a 32-bit unsigned value and returning the same value as a 64-bit unsigned integer) can be implemented as a no-operation at assembly level (with the special annotation, for the register allocator, that the target register should be the same as the source register).%
\footnote{This also explains why on some platforms, the code produced by {\CompCert} contains useless moves. If a 32-bit value needs to be extended to 64 bits in a way that both the 32-bit and 64-bit version are live after extension, then these two values, even if they are implemented by the same bit-string, will have to reside in two different registers, since {\CompCert} value semantics distinguishes 32-bit from 64-bit values.}
Similarly, if the application binary interface specifies that 32-bit values stored in 64-bit processor registers are sign-extended, then the sign-extension operation as defined in {\CompCert} can be implemented as a no-operation at assembly level.
Finally, the application binary interface may specify that the higher 32 bits of a 64-bit register containing a 32-bit value are arbitrary.

Since none of the {\CompCert} semantics specifies register contents at the bit level, it is up to the backend designer to be consistent in what instructions assume and ensure, and this consistency is never formally verified.
Consistency must extend to the foreign function interface: for instance, if a {\CompCert} function is called from a function compiled with another compiler that considers that the higher order 32 bits contain arbitrary values, but {\CompCert} assumes that values are $0$-extended, then incorrect behavior may ensue.

The modeling of certain instructions is delicate. The KVX processor supports, in addition to normal loads from memory, \emph{speculative} loads, otherwise known as \emph{non-trapping} or \emph{dismissible} loads. A normal load from an incorrect memory address will trap; on the KVX, a speculative load from an incorrect address returns $0$ instead of trapping. Here, ``incorrect'' is meant with respect to the page tables of the processor.%
\lstset{language=Coq}
In the intermediate representations of {\CompCertKVX}, speculative loads from incorrect memory locations return the special value ``undefined'', whereas a normal load would terminate execution.
``Undefined'' is a form of ``poison value'' propagating through operations, e.g., adding it to an integer yields ``undefined''.
The assembly-level semantics, however, defined the value returned by a speculative load from an incorrect memory location as~$0$, as per the processor documentation. $0$ is a valid refinement of ``undefined'', and the proofs go through.
This is however incorrect modeling, because it conflates two different notions: memory accesses invalid with respect to {\CompCert} semantics, and memory accesses invalid with respect to the processor memory management unit:%
\footnote{Or, rather, the association of the processor memory management unit and the virtual memory subsystem of the operating system.}
the former are strictly included in the latter:%
\footnote{In the case of memory over-commit by the OS, a valid memory access with respect to {\CompCert} semantics may result in a segmentation violation. We do not consider this issue here, since it is a case of the OS promising resources to the program then reneging on its promises, and thus not supplying a stable execution environment.},
a valid {\CompCert} memory block may occupy a portion of a valid memory page, but the processor will allow accesses to the whole page.
Using this incorrect semantics, one could perform a speculative load from a location known to be incorrect with respect to {\CompCert} semantics (for instance, just past the end of a block allocated on the stack) and assume that this load would return $0$, whereas this location, when read, would return another value.
Commit~\compcertkvxcommit{5798f56b}{5798f56b8a8630e43dbed84a824811a5626a1503} replaced this default value by ``undefined'', which is correct: any value is a valid refinement of~``undefined''.

\subsection{Foreign Function Interface}
{\CompCert}'s application binary interface (ABI) is not specified in a single point in {\CompCert}: it comprises the calling convention, the value conventions implicit in the choice of instructions, etc.
The correctness theorem of {\CompCert} relates the execution of a C program, started from the main function, to the execution of the assembly program produced by its compilation, also started from the main function.
It does not discuss functions compiled with other compilers calling a function compiled using {\CompCert}.
It also assumes that functions called from {\CompCert} use the same calling convention. As explained in {\CompCert}'s manual
\begin{quote}
{\CompCert} attempts to generate object code that respects the Application Binary Interface of the target
platform and that can, therefore, be linked with object code and libraries compiled by other C compilers.
\end{quote}
The manual then describes areas where {\CompCert}'s ABI differs from those of other compilers on the targets that it supports.
Again, none of these other ABIs were formalized, so the statement of differences in the manual is not based on formal analysis of compatibility, but rather on human analysis.

\subsection{Runtime System}
The runtime system for C is rather limited compared to other languages.
It uses the C standard library supplied by the target platform. {\CompCert} makes no assumption about it---calls to the standard library are just calls to external functions, and the sequence of these calls, as observable events, in the source semantics is reflected in the assembly code---except for the heap memory allocation and deallocation functions \lstset{language=C}\lstinline|malloc()| and \lstinline|free()|, which have special treatment and are given specific semantics (creation and destruction of memory blocks in the {\CompCert} memory model).
{\CompCert} assumes that this allocator is correct with respect to {\CompCert}'s \emph{infinite} memory model.
In particular, {\CompCert} assumes that \lstinline|malloc| always succeeds and never returns the null pointer, which seems unsound:
in theory, some formally verified optimizations may incorrectly remove defensive checks against heap overflow. 
In practice, we do not know of any optimization in {\CompCert} exploiting this model of \lstinline|malloc|. 
This assumption of infinite memory has been removed in {\CompCertS}\cite{DBLP:journals/jar/BessonBW19a}, at the price of a large extension of {\CompCert}. 

In {\CompCert}, basic floating-point operations have a semantics defined according to IEEE-754 in round-to-nearest mode. This assumes no change to the rounding mode through a library call or direct access to special CPU registers.

Some processors do not support some expensive arithmetic operations (e.g. floating-point operations, division) in hardware. These are replaced by calls to functions in the runtime system, which are axiomatized to perform the required operation by a combination of elementary instructions.
This creates a somewhat paradoxical situation where, for the same operation (say, 32-bit integer division):
\begin{inparaenum}[(i)]
\item if the operation is implemented in hardware, then it is trusted;
\item if implemented in software through a call to the runtime system, then it is trusted;
\item if implemented in software through expansion inside {\CompCert}, then one has to provide a full proof that this expansion implements the operation: its execution coincides with that of the operator on argument values on which this operator has defined behavior.
\end{inparaenum}
One argument is that the hardware is likely to have been designed from existant floating-point designs and thoroughly tested with many test vectors,%
\footnote{E.g. the Berkeley hard float library ({\scriptsize \url{https://github.com/ucb-bar/berkeley-hardfloat}}) is used in certain RISC-V designs. Yet, they remind potential users that ``These units are works in progress. They may not be yet completely free of bugs [...]''.}
Software emulation is likely to be from a well-tested established library,%
\footnote{E.g. the Berkeley soft float library ({\scriptsize \url{http://www.jhauser.us/arithmetic/SoftFloat.html}}); but, again ``Releases 3 through 3c of Berkeley SoftFloat contain bugs in the square root functions that may be of concern for some uses. Those bugs are believed to be repaired in Release 3d and later.''}
whereas expansion in {\CompCert} probably has not been tested so well.

\section{Insights and Conclusion}\label{sec:conclusion}


Some natural questions about ``verified'' software is: how truly safe is it? What kind of constructs should we be considered as suspicious?
As more designs come with some formal proofs of correctness, even regulatory agencies have had to provide guidelines~\cite{Coq_req_ANSSI_INRIA}.
It is of course perilous to draw general conclusions from the analysis of one single project; here are some insights. 

None of the problems found were in the verified parts of {\CompCert}:
chances seem slim to stumble into a proof checker bug by accident, not notice something is amiss, and think to have proved a theorem that actually does not hold.
This explains why the number of bugs found in {\CompCert} releases is many orders of magnitude below usual compilers~\cite{10.1145/2931037.2931074}.
By construction, the bugs of {\CompCert} are located in a limited subpart of the software, called its TCB, which may however not be as small as we may naively expect.

Two bugs were found in the front-end elaboration rules, ``corner cases'' that should be rarely found in real programs (thus their late discovery).
A few subtle semantic bugs were also found in some back-ends.
However, most bugs were found in the very last part of the back-end, which expands and prints assembly instructions.
The causes of these bugs are:
\begin{inparaenum}[(i)]
\item the tedium of writing correct printers for each instruction with appropriate operand ordering, and the lack of systematic unit testing of the printers;
\item the number of different cases, especially in the choice of register arguments, in the expansion of pseudo-instructions, and again the lack of systematic testing that all cases are correct;
\item the difficulties in keeping synchronized the specification of the pseudo-assembly instructions (in {\coq}) and the code performing their expansion, in two different files.
\end{inparaenum}
All these seem to be common software engineering issues, amenable to standard software engineering solutions such as systematic testing of all cases.

All these issues pertain to the specification and trusted (but unverified) parts of the {\CompCert} back-end, which echoes the results of early experiments that found bugs in these parts~\cite{YangCER11}.
In contrast, no bugs due to the use of axioms for interfacing untrusted code, or the use of the extractor to {\ocaml}, were found.
In academic circles, however, much attention is often given to doing away with such axioms and the extractor; this may not reflect the most pressing needs.
There seems to be a chasm between, on the one hand, what feels relevant and interesting for experts in proof assistants or type theoreticians, on the other hand what would actually increase reliability in verified compilers or similar tools.

In our opinion, the primary focus for increasing trust in {\CompCert} (and removing possible further bugs) should be a validation mechanism of its assembly and ABI specification with respect to the actual execution platform. For example, SAIL provides a formal ISA semantics for ARMv8 that has been tested against the ARM Architecture Validation
Suite~\cite{sail-popl2019}. However, {\CompCert} cannot be directly plugged on SAIL, because of its more abstract view of the ISA.
And this would not solve the issues related to the runtime environment and the ABI.

\section*{Acknowledgements}
We wish to thank A.~Miquel for helpful references on the metatheory of {\coq}, as well as L.~Gourdin, X.~Leroy and C.~Six for discussions about {\CompCert}.

\printbibliography
\end{document}